\documentclass[sigconf]{acmart}
\usepackage{caption}
\usepackage[justification=centering]{subcaption}
\usepackage{tikz-cd}
\usepackage{tikz}
\usetikzlibrary{quantikz2}
\usepackage{multirow}
\usepackage{balance}
\usepackage{enumitem}

\newcounter{RihanNOC}
\stepcounter{RihanNOC}

\newcounter{td}

\newcounter{all}

\newcommand{\para}[1]{\noindent\textbf{#1.}}

\newtheorem{exmp}{Example}[section]

\usepackage{xspace}
\newcommand{\sys}{Qymera\xspace}
\copyrightyear{2025}
\acmYear{2025}
\setcopyright{cc}
\setcctype{by}
\acmConference[SIGMOD-Companion '25]{Companion of the 2025 International Conference on Management of Data}{June 22--27, 2025}{Berlin, Germany}
\acmBooktitle{Companion of the 2025 International Conference on Management of Data (SIGMOD-Companion '25), June 22--27, 2025, Berlin, Germany}
\acmDOI{10.1145/3722212.3725126}
\acmISBN{979-8-4007-1564-8/2025/06}

\begin{document}

\title{\sys: Simulating Quantum Circuits using RDBMS}

 
\author{Tim Littau}
\affiliation{%
\institution{ 
Delft University of Technology}
\city{Delft}
\country{The Netherlands}
}
\email{t.m.littau@tudelft.nl}

\author{Rihan Hai}
\affiliation{%
\institution{ 
Delft University of Technology}
\city{Delft}
\country{The Netherlands}
}
\email{r.hai@tudelft.nl}

\renewcommand{\shortauthors}{Tim Littau and Rihan Hai}


\begin{abstract}

Quantum circuit simulation is crucial for quantum computing such as validating quantum algorithms. We present \sys, a system that repurposes relational database management systems (RDBMSs) for simulation by translating circuits into SQL queries, allowing quantum operations to run natively within an RDBMS. \sys supports 
a wide range of quantum circuits, offering a graphical circuit builder and code-based interfaces to input circuits. With a benchmarking framework, \sys facilitates comparison of RDBMS-based simulation against state-of-the-art simulation methods. Our demonstration showcases \sys's end-to-end SQL-based execution, seamless integration with classical workflows, and its utility for development, benchmarking, and education in quantum computing and data management. 
A video demonstrating Qymera's key features can be found at \url{https://youtu.be/j_fTKnVIV6c}.
\end{abstract}




\maketitle

\section{Introduction}
Quantum computing, heralded as the next frontier of computational science, promises revolutionary processing power beyond classical hardware. However, quantum computing remains in its early stages, known as the Noisy Intermediate-Scale Quantum (NISQ) era, where quantum hardware is limited by small numbers of qubits and noise. For instance,  Heron\footnote{\url{https://docs.quantum.ibm.com/guides/processor-types}}, one of IBM's most performant quantum processors, has only 156 qubits and lacks full error correction. In this landscape, \emph{simulation} serves as a core mechanism for exploring and validating quantum technologies, such as advancing quantum algorithm design, hardware development, and error correction \cite{xu2023herculeantaskclassicalsimulation}. 

Simulation refers to the use of classical computers to numerically reproduce the behavior of quantum circuits, tracking the evolution of quantum states and operations. 

\begin{figure}[t]
    \centering
    \includegraphics[width=0.45\textwidth]{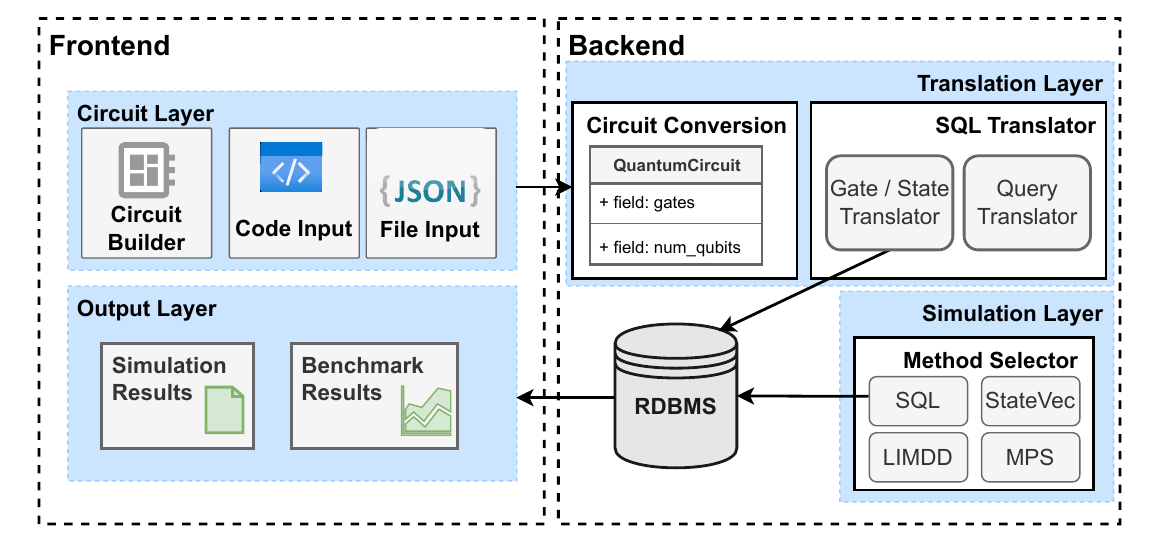}
    \caption{Overview of \sys}
    \label{fig:architecture}
     \vspace{-0.5cm} 
\end{figure}

Existing simulation frameworks such as Qiskit\footnote{Aer: \url{https://qiskit.github.io/qiskit-aer/}}, PyQuil\footnote{\url{https://pyquil-docs.rigetti.com/en/stable/}}, Cirq\footnote{qsim: \url{https://quantumai.google/qsim/cirq_interface}}, 
typically access data relying on programming languages, e.g., Python. They require users to specify the simulation methods.   
To improve performance, low-level optimizations are needed, such as parallelization and Single-Instruction, Multiple-Data (SIMD) \cite{young2023simulatingquantumcomputationsclassical}.  This approach can be cumbersome, especially for large-scale simulations, and places a high burden on developers to fine-tune performance.

\begin{figure*}[ht]
    \centering
    \includegraphics[width=1.0\textwidth]{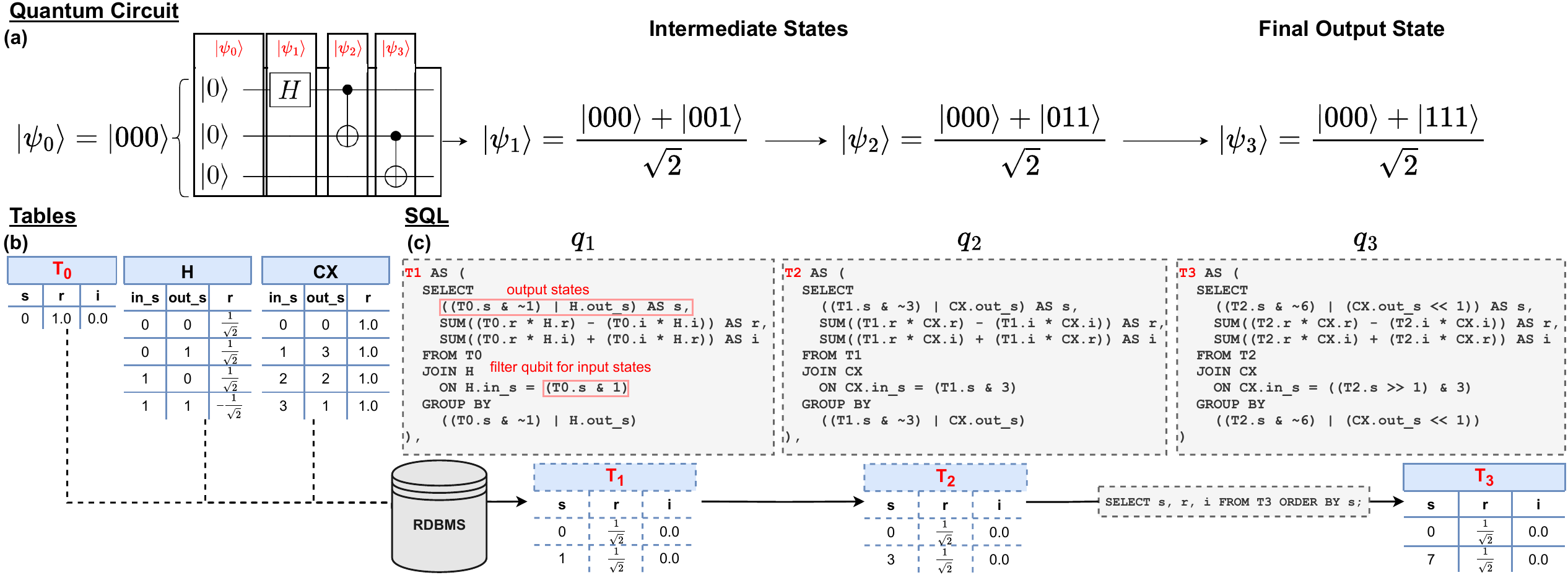}
\caption{Running example of the SQL translation flow: 
(a) the input circuit transforming initial state \(\ket{\psi}_0\) into final state \(\ket{\psi}_3\), where horizontal wires represent qubits, and boxes indicate initial, intermediate, and final states; 
(b) tables for the initial state \(\ket{\psi}_0\), the H gate, and the CNOT gate; and 
(c) SQL queries \(\{q_1, q_2, q_3\}\) generated for each gate operation in the input circuit, along with the resulting intermediate tables \(T_1, T_2\) and final output table \(T_3\).}
        
    \label{fig:sqlflow}
      \vspace{-0.4cm}
\end{figure*}
\para{Our proposal}
We introduce a novel method that \emph{utilizes relational database management systems (RDBMSs) for simulating quantum circuits}. 
In this approach, quantum circuits—comprising a series of quantum gates applied to qubits—are represented by storing qubit and gate information in tables. 
The simulation process is then driven by SQL queries that specify the required computations. By leveraging the declarative nature of RDBMSs, simulation engineers and database researchers can concentrate on defining the desired outcomes without delving into the underlying execution details. RDBMSs then handle the complexity by automatically optimizing and executing these queries through techniques such as logical and physical query planning, caching,  and hardware-aware techniques such as parallelism~\cite{schmidt12rethinking}.

However, using RDBMSs as an out-of-the-box solution for simulation poses quantum-specific challenges. First, quantum circuits often involve bit-level operations (e.g., manipulating specific qubits), which SQL does not natively support. Second, although RDBMS-based simulation can outperform existing simulation methods for certain circuits, they are not universally optimal \cite{hai2025quantum}. In our preliminary experiments with a 2.0 GB memory limit, RDBMS approach simulated up to 3,118× more qubits than a conventional simulation method for sparse circuits but performed 14\% worse on dense circuits (see appendix B4, Fig. 10 in \cite{techreport}). Consequently, it is critical to identify scenarios where RDBMSs excel by benchmarking them against existing simulation methods, rather than applying them blindly.

We present \emph{\sys}, a novel system for simulating quantum circuits using RDBMSs. \emph{\sys} translates quantum circuits into SQL queries, enabling RDBMS-based simulations.
Its key features include: 1) a graphical circuit builder for intuitive design and parameterized circuits via Qiskit- or PyQuil-like syntax, 2) a unique feature of supporting RDBMS-based simulation,  alongside state-of-the-art simulation methods such as state-vector, tensor networks, e.g., MPS, and LIMDD.\footnote{These simulation methods (also known as data representations) are a complex and active research area in quantum computing. We provide detailed definitions, explanations and examples in Appendix A and B1 of our online technical report \cite{techreport}.} and 3) a benchmarking suite for systematically comparing RDBMS performance against alternative simulators on a wide range of circuit inputs. 

\section{Transforming Quantum Circuits into SQL}
\label{sec:transforming-quantum-circuits}

A quantum circuit consists of a sequence of gates, which operate on quantum states. 
Simulating quantum circuits involves translating the abstract mathematical representation of quantum states and gates into computational operations. 

In this section, we explain how \sys maps quantum states and gates to tables and translates circuit operations into SQL queries while addressing quantum-specific challenges. We walk through the SQL translation process using a running example in Fig.~\ref{fig:sqlflow}.

\subsection{From Quantum States and Gates to Tables}  
An $n$-qubit quantum state is represented as  
\[
\vspace{-0.1cm}
\ket{\psi}= \sum_{j=0}^{2^n - 1} \bigl(a_j + i\,b_j\bigr) \ket{j},
\]  
where \( j \) is an integer index ranging from \( 0 \) to \( 2^n - 1 \), $\ket{j}$ is the computational basis state (binary strings of length $n$, such as $\ket{0\cdots 0}$ to $\ket{1\cdots 1}$). 
Each basis state \( \ket{j} \) has a complex amplitude \( a_j + i b_j \), where \( a_j \) and \( b_j \) are the real and imaginary parts, respectively.  

We define the relational schema for a state \(\ket{\psi}\) as \texttt{T(s, r, i)}, where \( s \) is the computational basis state $\ket{j}$ encoded as an integer, and \( r \) and \( i \) represent the real and imaginary parts of its complex amplitude, both being real-valued. Only nonzero basis states are stored.

A quantum gate transforms an input quantum state into an output state. We represent this transformation in a relational table that captures the integer indices of the input and output states, along with their transition amplitudes. We define the relational schema for a gate as \texttt{T(in\_s, out\_s, r, i)}, where \texttt{in\_s} and \texttt{out\_s} denote input and output state indices, and \texttt{r} and \texttt{i} represent the real and imaginary parts of the transition amplitude.

\begin{exmp}[Quantum states and gates as tables]
\label{exmp:ghz}
Consider the circuit shown in Fig.~\ref{fig:sqlflow}a), generating a 3-qubit GHZ state. Initially, the state is \(\ket{000}\), represented as \texttt{(0, 1.0, 0)}. A Hadamard gate mapping input indices to outputs, each with amplitude \(\tfrac{1}{\sqrt{2}}\) applied to the first qubit transforms it into the superposition \(\frac{1}{\sqrt{2}} (\ket{000} + \ket{100})\), stored in the relational form as tuples \texttt{(0, $\frac{1}{\sqrt{2}}$, 0)} and \texttt{(1, $\frac{1}{\sqrt{2}}$, 0)}. Subsequent CX gates manipulate the state further by updating the table to represent resulting intermediate states \(\ket{\psi}_2\) and \(\ket{\psi}_3\).

\end{exmp}

\subsection{From Simulation Operations to SQL}
Simulating quantum circuits requires bit-level manipulation, since each gate operates on specific qubits. To enable this in SQL, we adopt the previously explained \emph{integer encoding} and apply \emph{bitwise operations} (see Table~\ref{tab:bitwise_operations}) to pinpoint the relevant qubit and perform bit-level gate operations.

\begin{exmp}[H Gate Operation via SQL]
\label{exmp:h2sql}
\vspace{-0.1cm}

Query \( q_1 \) in Fig.~\ref{fig:sqlflow}c) represents the Hadamard gate operation from Fig.~\ref{fig:sqlflow}a). Since the gate acts on the first qubit, we locate it in \texttt{T0} using \texttt{T0.s \& \textasciitilde 1} and join with \texttt{H} table via  
\texttt{JOIN H ON H.in\_s = (T0.s \& 1)}.  
The value of \texttt{s} of the output state is computed as \texttt{(T0.s \& \textasciitilde 1) | H.out\_s}. 
\end{exmp}

 We can see from the above example that by using bitwise operators in  Table~\ref{tab:bitwise_operations}, we can directly locate and manipulate individual qubits and directly access and operate on them.

\para{Discussion} We have presented a declarative, SQL-based approach that specifies \emph{what} operations gates perform without requiring users to manage \emph{how} to implement them. Our work improves existing methods \cite{sqleinsum, Trummer24} in the following aspects. First, Blacher et al. \cite{sqleinsum} encode gates as tensor computations, often requiring multiple columns per index dimension, leading to no clear performance advantage over NumPy, as shown in their experiments. By contrast, our integer-based encoding leverages bitwise operations, reducing storage overhead and potentially improving performance.  \cite{Trummer24} treats qubit states as strings, which increases storage costs and complicates indexing. 
Our approach includes CPU-native bitwise instructions and the integer indexing of qubit state, enabling more efficient lookups and compact storage than string-based representations in \cite{Trummer24}.

\section{System Overview}
\label{sec:system-overview}

\sys is an end-to-end, web-based system demonstrating how RDBMSs are leveraged for quantum circuit  simulation. As depicted in Fig.~\ref{fig:architecture}, it organizes its functionality across four different layers. 

\subsection{Circuit Layer}
\label{subsec:circuit-layer}

Existing frameworks like Qiskit and Cirq offer powerful APIs but lack intuitive graphical interfaces, limiting accessibility for non-programmers. Browser-based tools like Quirk\footnote{\url{https://algassert.com/quirk}} and the Quantum Länd Simulator\footnote{\url{https://thequantumlaend.de/quantum-circuit-designer/}} attempt to bridge this gap but have significant limitations: Quirk, built on JavaScript, supports only 16 qubits and lacks parameterized circuit support, while the Quantum Länd Simulator provides a QASM interface with limited and unintuitive handling of parameterized circuits. 

To address these shortcomings, we introduce the \emph{Circuit Layer}, shown in Fig.~\ref{fig:CircuitUI}, a flexible, user-friendly interface for quantum circuit construction and integration.
It provides multiple methods for circuit input, including:
\begin{itemize}[leftmargin=*]
    \item \textbf{Graphical Circuit Builder:} An intuitive drag-and-drop interface that allows researchers, particularly those new to quantum computing, to visually construct quantum circuits. 
    \item \textbf{File Upload:} Quantum researchers can upload circuits in standardized formats, such as JSON, to import pre-defined designs.
    \item \textbf{Parameterized Circuit Families:} Researchers can define parameterized circuits programmatically using various APIs. 
\end{itemize}

\begin{table}[t]
\centering
\begin{tabular}{| c | c | l |}
\hline
\textbf{Bitwise Operation} & \textbf{Symbol} & \textbf{Description} \\
\hline
AND & \texttt{\&} & Bitwise AND of two operands\\
\hline
OR  & \texttt{|} & Bitwise OR of two operands \\
\hline
NOT & \texttt{\textasciitilde} & Inverts bitstring of operand\\
\hline
Left Shift  & $<<$ & Shifts bits left by $n$ bits. \\
\hline
Right Shift & $>>$ & Shifts bits right by $n$ bits. \\
\hline
\end{tabular}
\caption{Bitwise Operations used in the SQL queries}
\vspace{-1cm}
\label{tab:bitwise_operations}
\end{table}

\subsection{Translation Layer}
\label{subsec:translation-layer}

The \emph{Translation Layer}, as detailed in Sec.~\ref{sec:transforming-quantum-circuits}, is responsible for converting abstract quantum circuits into executable SQL queries. 
\begin{itemize}[leftmargin=*]
    \item \textbf{SQL Generation:} The translation layer parses the input circuit, identifies qubits and gates involved, and generates the corresponding SQL queries.
    \item \textbf{Gate Representation:} Quantum gates  are stored as   tables encoding input-output mappings and transition probabilities.
    \item \textbf{Query Optimization:} To improve performance, consecutive gates are fused into single SQL query 
    where possible, minimizing intermediate results and leveraging database query optimizers.
\end{itemize}

\subsection{Simulation Layer}
\label{subsec:simulation-layer}

The \emph{Simulation Layer} handles the execution of the SQL queries generated in the Translation Layer. Quantum researchers can manage simulation runs as shown in Fig.~\ref{fig:SimUI}.

\begin{itemize}[leftmargin=*]
    \item \textbf{Backend Integration:} \sys executes SQL queries on RDBMS engines to compute the resulting quantum state or measurement probabilities. It supports SQLite 2.6.0, and DuckDB 1.1.
    \item \textbf{Support for Multiple Methods:} \sys allows quantum and database researchers to compare database-driven simulations with alternative simulation methods like tensor network and state-vector simulators (cuQuantum \cite{cuQuantum}) or decision diagram based simulators (MQT DD \cite{mqtdd}).
    \item \textbf{Parameterized Simulations:} Quantum researchers can define families of circuits with varying parameters, and \sys automates simulation across the parameter space.
    \item \textbf{Out-of-Core Simulation:} For large circuits that exceed memory capacity, \sys leverages database features to efficiently manage intermediate states and I/O, enabling simulations at scales beyond traditional in-memory methods.
\end{itemize}

\subsection{Output Layer}
\label{subsec:output-layer}

\begin{figure}[t]
\begin{subfigure}[b]{0.5\textwidth}
    \centering
    \includegraphics[width=0.9\textwidth]{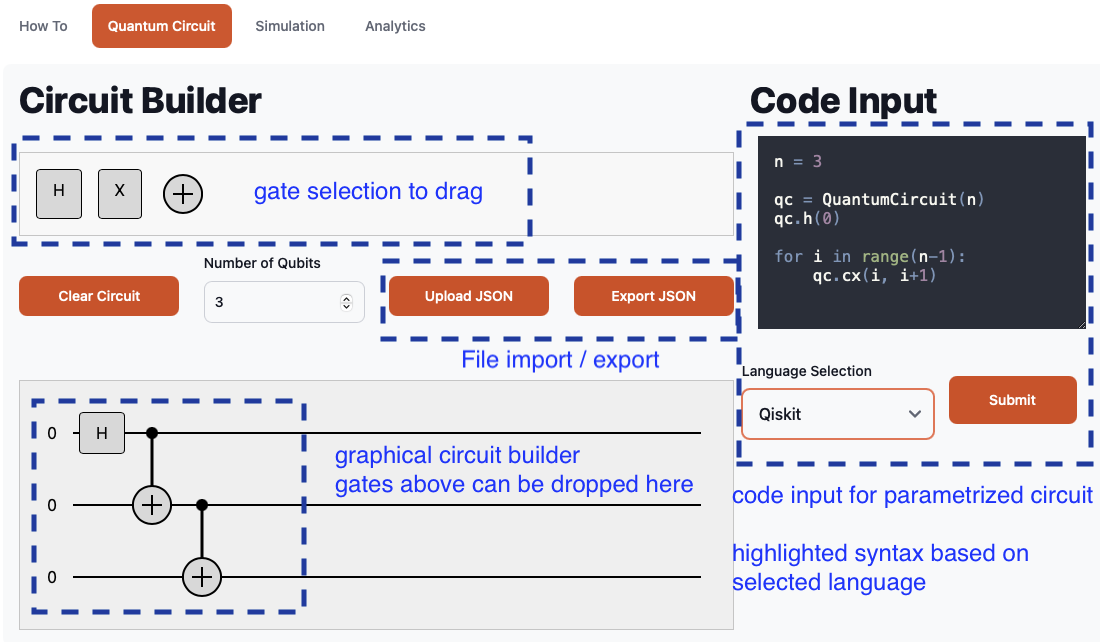}
    \subcaption{Circuit Panel Tab}
    \label{fig:CircuitUI}
\end{subfigure}
\begin{subfigure}[b]{0.5\textwidth}
    \centering
    \includegraphics[width=0.9\textwidth]{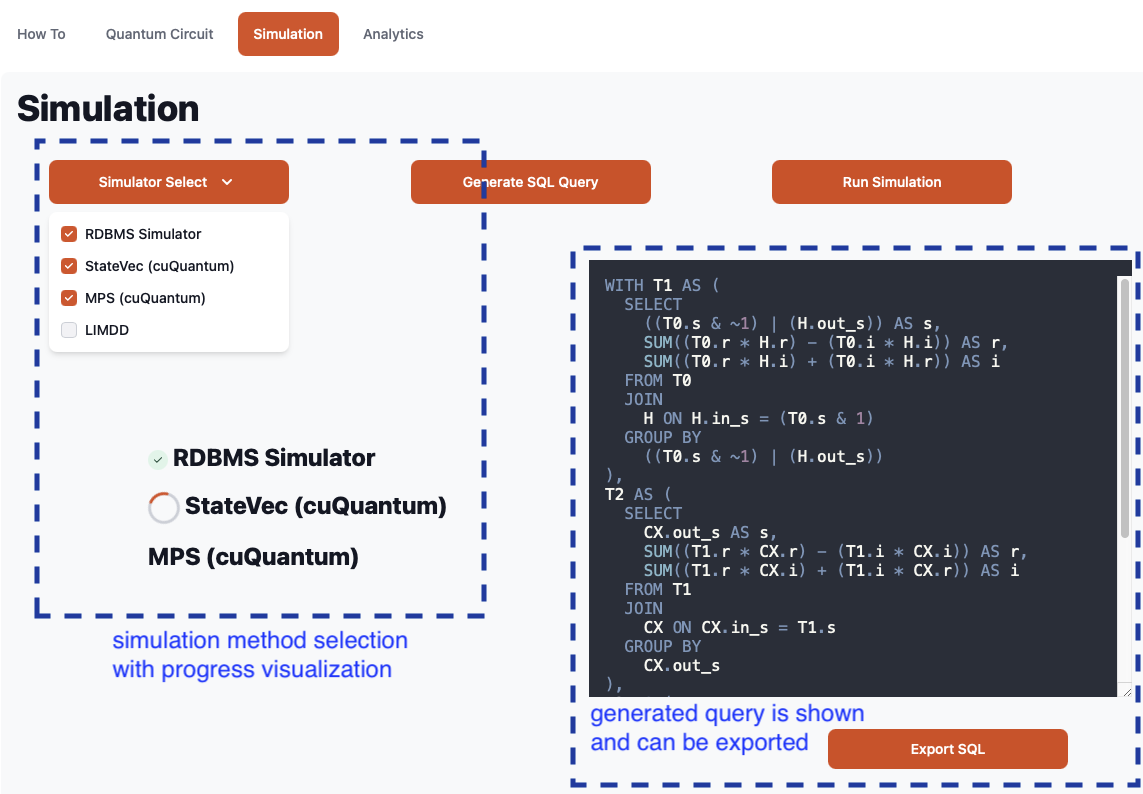}
    \subcaption{Simulation Panel Tab}
    \label{fig:SimUI}
\end{subfigure}
\begin{subfigure}[b]{0.5\textwidth}
    \centering
    \includegraphics[width=0.9\textwidth]{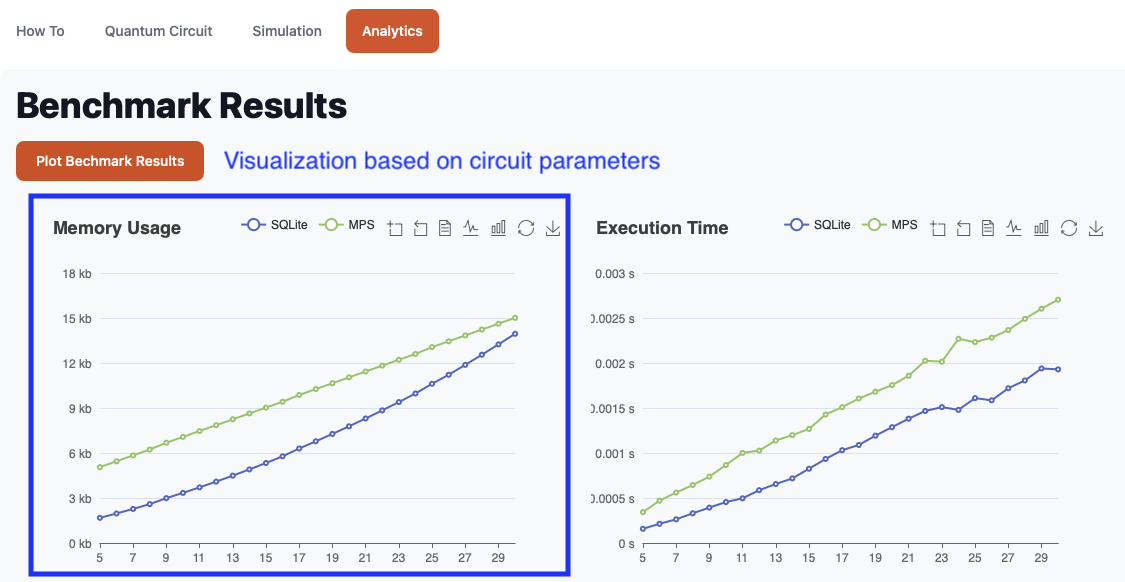}
    \subcaption{Visualization Panel Tab}
    \label{fig:OutputUI}
\end{subfigure}
\caption{\sys's core UI tabs.}
\label{fig:ui}
\vspace{-0.3cm}
\end{figure}

The \emph{Output Layer}, as illustrated in Fig.~\ref{fig:OutputUI},
provides researchers with tools to analyze simulation results and benchmarking metrics. 

\begin{itemize}[leftmargin=*]
    \item \textbf{Simulation Results:} The system outputs the final quantum state, including measurement probabilities.
    \item \textbf{Performance Metrics:} Execution time, memory usage, and other relevant metrics are logged and displayed for each simulation method.
    \item \textbf{Visualization Tools:} Interactive plots enable the user to explore the behavior of circuits, compare simulation methods, and analyze performance across parameterized families of circuits.
    \item \textbf{Export and Reporting:} Results and visualizations can be exported for analysis or publication.
\end{itemize}
The Output Layer ensures that researchers derive insights from their simulations, through detailed analysis and high-level comparisons.

\section{Demonstration Scenarios}
\label{sec:demonstration-scenarios}

The demonstration of \sys will showcase its application in quantum circuit simulation using RDBMSs, serving both as a simulation method and an infrastructural backbone for circuit exploration. SIGMOD attendees will interact with the system through an intuitive interface as shown in Fig.~\ref{fig:ui}. They can explore \sys's ability to convert quantum circuits into SQL queries, execute them efficiently within an RDBMS, and benchmark parameterized quantum circuits using various simulation backends. 

The demonstration is structured into the following scenarios.
\balance
\para{Quantum Algorithm Design and Testing} This scenario highlights how \sys enables rapid iteration and testing of quantum algorithms. For example, attendees will construct and analyze a simple quantum parity check algorithm, which determines whether the number of ones in a given bitstring is even or odd. 

\sys will translate the algorithm into SQL, execute it within an RDBMS, and allow attendees to inspect intermediate quantum states and measure execution performance. Additionally, attendees can compare the execution with other simulation techniques, such as state-vector simulation, to assess the trade-offs in performance and accuracy. This will demonstrate how researchers and developers can leverage \sys to refine quantum algorithms efficiently and select the optimal simulation method for their needs.
\para{Simulation Method Benchmarking}
Benchmarking different quantum circuit simulators is crucial for assessing performance trade-offs. In this scenario, attendees will evaluate the efficiency of various simulation backends using a GHZ state preparation circuit and an equal superposition of all possible states as test cases. The system will execute the circuit across different simulation approaches as introduced in Sec.~\ref{sec:system-overview}.

Performance metrics, including execution time, and memory usage, will be analyzed, providing insights into the strengths and limitations of each simulator. This comparative analysis will highlight the conditions under which SQL-based simulation is advantageous and when alternative simulators may be preferable.

\para{Educational Exploration of Quantum Computing Concepts}
\sys also serves as a tool to explore fundamental quantum computing principles. This scenario will focus on illustrating entanglement and superposition using the GHZ state as a case study. Attendees will construct a circuit, observe the evolution of quantum states through SQL queries, and explore measurement outcomes. 

By interacting with \sys, SIGMOD attendees will develop a deeper understanding of quantum mechanics concepts. 

To further enhance educational effectiveness, \sys will integrate interactive visualizations explicitly focused on quantum states and gate operations. Specifically quantum states visually represented as Block spheres and how they evolve upon applying different gates in the graphical circuit builder. This scenario is designed to provide a conceptual and interactive learning experience, making quantum computing principles more accessible. 

Through the above scenarios, participants will gain a comprehensive understanding of \sys, including its novel SQL-based approach to quantum circuit simulation, its role in benchmarking different simulation methods, and its educational potential.

\begin{acks}
This publication was supported (in part) by Dutch Research Council (VI.Veni.222.439).
\end{acks}
\vspace{0.27cm}
\bibliographystyle{ACM-Reference-Format}

\bibliography{ref}


\end{document}